\documentclass{webofc}
\usepackage[varg]{txfonts}   % Web of Conferences font
\usepackage{graphicx,amsmath,amssymb}
\usepackage{hyperref}
\usepackage{color}
\usepackage{lineno}

\definecolor{navyblue}{rgb}{0,0.08,0.45}
\definecolor{darkred}{rgb}{0.7,0.0,0.0}
\definecolor{darkgreen}{rgb}{0,0.6,0.2}

\newcommand{\beq}{\begin{equation}}
\newcommand{\enq}{\end{equation}}
\newcommand{\beqa}{\begin{eqnarray}}
\newcommand{\beqast}{\begin{eqnarray*}}
\newcommand{\enqa}{\end{eqnarray}}
\newcommand{\enqast}{\end{eqnarray*}}

\newcommand{\req}[1]{(\ref{#1})}

\newcommand{\half}{{\frac{1}{2}}}

\newcommand{\bec}{\begin{center}}
\newcommand{\enc}{\end{center}}
\newcommand{\beqo}{\begin{quote}}
\newcommand{\enqo}{\end{quote}}

\newcommand{\ze}{\zeta}

\newcommand{\la}{\lambda}

\newcommand{\si}{\sigma}

\newcommand{\ph}{\phi}

\newcommand{\vp}{\varphi}

\newcommand{\La}{\Lambda}

\begin{document}
\title{Superconformal Algebraic Approach to Hadron Structure}

\author{\firstname{Guy F.} \lastname{de T\'eramond}\inst{1}\fnsep\thanks{\email{gdt@asterix.crnet.cr}} 
\and
        \firstname{Stanley J.} \lastname{Brodsky}\inst{2}\fnsep\thanks{\email{sjbth@slac.stanford.edu}} 
\and
        \firstname{Alexandre} \lastname{Deur}\inst{3}\fnsep\thanks{\email{deurpam@jlab.org}}
\and
        \firstname{Hans G\"{u}nter} \lastname{Dosch}\inst{4}\fnsep\thanks{\email{h.g.dosch@thphys.uni-heidelberg.de}}
\and
        \firstname{Raza Sabbir} \lastname{Sufian}\inst{5}\fnsep\thanks{\email{sabbir.sufian@uky.edu}}               
}

\institute{Universidad de Costa Rica, 11501 San Jos\'e, Costa Rica
\and
           SLAC National Accelerator Laboratory, Stanford University, Stanford, CA 94309, USA
\and
          Thomas Jefferson National Accelerator Facility, Newport News, VA 23606, USA 
\and 
          Institut f\"{u}r Theoretische Physik der Universit\"{a}t, Philosophenweg 16, D-69120 Heidelberg, Germany
\and
         Department of Physics and Astronomy, University of Kentucky, Lexington, KY 40506, USA}

\abstract{Fundamental aspects of nonperturbative QCD dynamics which are not obvious from its classical Lagrangian, such as the emergence of a mass scale and confinement, the existence of a zero mass bound state,  the appearance of universal Regge trajectories  and the breaking of chiral symmetry are incorporated from the onset in an effective theory based on superconformal quantum mechanics and its embedding in a higher dimensional gravitational theory. In addition, superconformal quantum mechanics gives remarkable connections between the light meson and nucleon spectra. This new approach to hadron physics is also suitable to describe nonperturbative QCD observables based on structure functions, such as GPDs, which are not amenable to a first-principle computation. The formalism is also successful in the description of form factors, the nonperturbative behavior of the strong coupling and diffractive processes. We also discuss in this article how the framework can be extended rather successfully to the heavy-light hadron sector.}
\maketitle
\section{Introduction}
\label{intro}

The fundamental interactions of quarks and gluons are remarkably well described by QCD in the short-distance perturbative domain where it has been subject to high precision experimental test. However, the increase of the QCD coupling at low energies implies that an infinite number of quark and gluons are strongly coupled and the description of the dynamics becomes a vastly complex problem. Indeed  understanding the mechanism of confinement is so far an unsolved problem~\cite{Brambilla:2014jmp}, notwithstanding that the QCD Lagrangian is well established.  Euclidean lattice gauge theory is a first-principle numerical simulation of nonperturbative QCD. Still, lattice computation of the resonance spectrum of the light hadrons, and particularly baryons, represents a formidable task due to the computational complexity beyond the leading ground state configuration~\cite{Edwards:2011jj}. In addition, important QCD observables such as Generalized Parton Distributions (GPDs), which are most relevant for the analysis of actual experiments  --such as experiments at the LHC and JLab, are presently not amenable to a first-principle computation. 

In a recent series of articles~\cite{deTeramond:2014asa, Dosch:2015nwa, Dosch:2015bca, Brodsky:2016yod} we have shown  how to use superconformal quantum mechanics~\cite{Fubini:1984hf} to construct relativistic light-front semiclassical bound-state equations, which can be embedded in a higher dimensional classical gravitational theory. This new approach to hadron physics incorporates basic aspects of nonperturbative dynamics which are not obvious from the QCD Lagrangian: The emergence of a mass scale and confinement out of a classically scale invariant theory as well as the appearence of a zero mass particle, the pion, in the limit of zero quark masses,  universal Regge trajectories  and the breaking of chirality in the hadron excitation spectrum. Furthermore, this new approach to hadronic physics gives remarkable connections between the light meson and nucleon spectra~\cite{Dosch:2015nwa} as well as predictions for the heavy-light hadron spectra, where heavy quark masses break the conformal invariance, but the underlying dynamical supersymmetry still holds~\cite{Dosch:2015bca}.

In addition to providing a relativistic frame-independent first-approximation to the light-front (LF) effective Hamiltonian which reproduces  basic QCD features, the embedding of the superconformal algebraic construction in a modified anti-de Sitter (AdS)  five-dimensional space provides us with critical nonperturbative elements required for the description of processes in QCD in particular domains: For example, the incorporation of the nonperturbative pole structure in the computation of hadronic electromagnetic form factors~\cite{Brodsky:2014yha}. Furthermore, the holographic embedding also incorporates the high-energy power-counting scaling behavior~\cite{Brodsky:1973kr, Matveev:1973ra, Polchinski:2001tt}, including asymptotic predictions, thus also providing a description of the transition from nonperturbative to perturbative domains. This interface of soft and hard regimes has also been described for the QCD coupling~\cite{Deur:2016cxb}  and for holographic models of  structure functions, such as GPDs~\cite{Gutsche:2013zia, Gutsche:2016gcd, Bacchetta:2016wby, Maji:2016yqo, Chakrabarti:2013dda, Traini:2016jko}. In the latter case, the structure functions are modeled in the hadronic domain from the LF holographic mapping up to a transition scale, where perturbative QCD controls further evolution to high energy scales. The light-front holographic framework has  also been used to describe other processes such as double parton~\cite{Traini:2016jru} and diffractive scattering~\cite{Forshaw:2012im}, as well as $B$ form factor decays~\cite{Ahmady:2013cga}.

\section{The Superconformal Light-Front Hamiltonian}
\label{SCH}

The superconformal light-front Hamiltonian $H$ can be expressed in the formalism of supersymmetric quantum mechanics~\cite{Witten:1981nf}  in terms of two fermionic generators, the supercharges, $Q$ and $Q^\dagger$ with the anticommutation relations 
\beq 
\{Q,Q\} = \{Q^\dagger,Q^\dagger\}=0,   
\enq
and the Hamiltonian $H$
\beq   \label{H}
H=  \{Q,Q^\dagger\},
\enq
which commutes with the fermionic generators ${[Q, H]}  = [Q^\dagger, H] = 0$, closing the graded Lie algebra. Since the Hamiltonian $H$ commutes with $Q^\dagger$, it  follows that the states  $\vert \phi \rangle$ and $Q^\dagger  \vert \phi \rangle$ have identical non-vanishing eigenvalues. Furthermore, if $|\ph_0 \rangle$ is an eigenstate of $Q$ with zero eigenvalue, it is annihilated by the operator $Q^\dagger$: $Q^\dagger|\ph_0 \rangle = 0$. Consequently, the lowest state, here identified as the pion, is not paired~\cite{Witten:1981nf, Dosch:2015nwa}. Thus the special role of the pion in the supersymmetric approach to hadronic physics as a unique state of zero energy in the limit of massless quarks.

In matrix notation 
\beq 
Q = \left(\begin{array}{cc}
0 & q\\
0 & 0\\
\end{array}
\right) , \quad  \quad
Q^\dagger=\left(\begin{array}{cc}
0 & 0\\
q^\dagger & 0\\
\end{array} \right),  \quad \quad
H=  
\left(\begin{array}{cc}
q \, q^\dagger &  0\\
0 & q^\dagger q \\
\end{array}
\right) 
\enq
with
\beq \label{qqdag} 
q = -\frac{d}{d \ze} + \frac{f}{\ze} + V(\ze),  \quad \quad
q^\dagger = \frac{d}{d\ze}  + \frac{f}{\ze} + V(\ze),
\enq
where $\ze$ is the invariant transverse separation between constituents\footnote{The invariant light-front variable of an $N-$quark bound state is $\zeta=\sqrt{\frac{x}{1-x}}\big\vert\sum_{j=1}^{N-1}x_j\bf{b}_\perp\big\vert$, where $x$ is the longitudinal momentum fraction of the active quark, $x_j$ with $j =1, 2, \cdots, N-1$, the momentum fractions associated with the $N-1$ quarks in the cluster, and ${\bf{b}}_{\perp j}$ are the transverse positions of the spectator quarks in the cluster relative to the active one~\cite{Brodsky:2006uqa}. For a two-constituent bound-state $\zeta =\sqrt{x(1-x)}\vert{\bf{b}}_\perp\vert$.} in the light-front quantization scheme, which is identified with the holographic variable $z$ in the AdS classical gravity theory~\cite{Brodsky:2006uqa}, and  $f$ is a dimensionless constant.

If the superpotential $V$ is vanishing, the  Hamiltonian is also invariant under conformal transformations and one can extend the supersymmetric algebra to a  superconformal algebra~\cite{Fubini:1984hf, Akulov:1984uh}.  Following the procedure introduced in Refs.~\cite {deAlfaro:1976je, Brodsky:2013ar}  the Hamiltonian is constructed as a superposition of the superconformal generators which carry different dimensions, leading to the natural emergence of a scale, while the action remains conformally invariant.  For supersymmetric quantum mechanics this procedure determines uniquely the superconformal potential in \req{qqdag}: It is given by $V= \sqrt{\la} \,\ze$~\cite{deTeramond:2014asa, Dosch:2015nwa, Fubini:1984hf}. The LF Hamiltonian equation $H \vert \phi \rangle = M^2 \vert \phi \rangle$ acts on the spinor 
$\vert \phi \rangle =
\left( \begin{array}{c} \phi_M\\
\phi_B \end{array} \right)$\, leading to the  equations~\cite{Dosch:2015nwa}
\beqa 
\left(- \frac{d^2}{d \ze^2}+\frac{4 L_M^2-1}{4\ze^2}+U_M(\ze) \right) \phi_M &\!=\!& M^2 \phi_M, \label{HM}\\
\left(- \frac{d^2}{d \ze ^2} +\frac{4 L_B^2-1}{4\ze ^2}+U_B(\ze ) \right) \phi_B &\!=\!& M^2 \phi_B, \label{HB}
\enqa
where $L_B + \half = L_M - \half = f$ and the LF potential $U_M$ and $U_B$ are, respectively, the meson and baryon LF confinement potentials:
\beqa  
U_M(\ze) &\!=\!&    V^2(\ze) - V'(\ze) + \frac{2L_M - 1}{\ze} V(\ze) \, \to \, \la_M^2 \ze^2 + 2 \, \la_M (L_M - 1),   \label{UM} \\
U_B(\ze) &\!=\!&    V^2(\ze) + V'(\ze) + \frac{2L_B+1}{\ze} V(\ze) \, \to \, \la_B^2 \ze^2 +  2 \, \la_B (L_B + 1),   \label{UB}
\enqa    
where the right hand side in Eqs. (\ref{UM}, \ref{UB}) is the conformal limit for $V = \sqrt{\la} \, \ze$ with $\la_M = \la_B = \la$.

The superconformal framework also leads to relativistic bound-state equations for nucleons from the mapping to light-front physics and uniquely determines the confinement potential~\cite{deTeramond:2014asa}. In this case the LF Hamiltonian acts on the spinor $\vert \psi \rangle =
\left( \begin{array}{c} \psi_+\\
\psi_- \end{array} \right)$\, leading to the equations~\cite{deTeramond:2014asa}
\beqa 
\left(- \frac{d^2}{d \ze^2}+\frac{4 L^2-1}{4\ze^2} + \la^2 \ze^2 + 2 \la (L + 1) \right) \psi_+ &\!\!=\!\!& M^2 \psi_+, \label{psip} \\
\left(- \frac{d^2}{d \ze ^2} +\frac{4 (L+1)^2-1}{4\ze ^2} + \la^2 \ze^2 +  2  \la  L   \right) \psi_- &\!\!=\!\!& M^2 \psi_-, \label{psim}
\enqa
where the upper component $\psi_+$ is the leading twist positive chirality solution with orbital angular momentum $L$, whereas $\psi_-$ is the non-leading twist minus chirality component with angular momentum $L + 1$. Comparing  \req{psip} with \req{HB} and \req{UB}  it becomes clear that the lower component $\phi_B$ in \req{HB}  corresponds to the upper component $\psi_+$ in \req{psip}~\cite{deTeramond:2014asa, Brodsky:2014yha} with angular momentum $L= L_B=L_M -1$. On the other hand, in LF holographic QCD  the confinement potential for  mesons  $U_M$ is introduced by a dilaton term $e^{\vp(z)}$ in the AdS$_5$ action~\cite{Karch:2006pv}. It is given by~\cite{deTeramond:2010ge}
\beq \label{dilU} 
 U_M(\zeta)=  \frac{1}{4}(\vp'(\ze))^2  + \frac{1}{2} \vp''(\ze) + \frac{2 J -3}{2 \ze} \vp'(\ze) ,
\enq
and leads, for $J = L_M$, to the LF potential \req{UM} if we choose the dilaton profile $\varphi = \la z^2$.  We thus identify the upper component $\phi_M$ in \req{HM} with a meson with angular momentum $L_M$, with identical mass $M^2$ as its partner nucleon with orbital angular momentum $L_B = L_M -1$. The superconformal framework described here also incorporates a doublet consisting of the non-leading twist minus chirality component $\psi^-$, Eq. \req{psim}, of a baryon with angular momentum $L_B + 1$ and its partner tetraquark with angular momentum $L_T = L_B$~\cite{Brodsky:2016yod}.  The tetraquark sector is discussed in more detail in Ref.  \cite{Brodsky:2016yod}.  

\begin{figure*}
\centering
\sidecaption
\includegraphics[width=6.4cm]{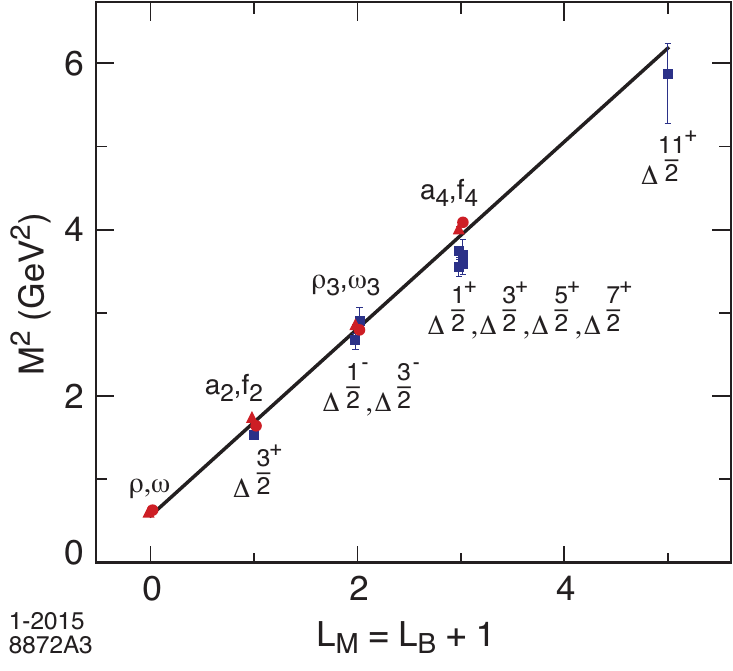} 
\caption{Superconformal vector meson-$\Delta$ partners: solid  line corresponds to $\sqrt \la= 0.53$  GeV. The data is from Ref. \cite{Olive:2016xmw}.}
\label{VMD}     
\end{figure*}

So far we have not considered the spin-spin interaction. For a spin-$J$ meson the spin interaction is computed straightforwardly from \req{dilU}. It leads to the additional interaction term  $2 \la \,  s$, where $s$ is the internal quark spin of the meson, which is also the quark spin of the diquark cluster in the baryon if we retain the meson-baryon supersymmetry. Thus the modified Hamiltonian~\cite{Brodsky:2016yod}
\beq   \label{Hs}
H=  \{Q,Q^\dagger\} + 2 \la s .
\enq
The effect of the spin interaction term is an overall shift of the energy levels, thus all our previous considerations remain the same, in particular the fact that the lowest meson for each trajectory has not a baryon partner, a property which is observed all across the hadron spectrum. As an illustration we compare in  Fig. \ref{VMD}  the measured values of the unflavored vector mesons and  $\Delta$ baryons with the predictions from the superconformal model described here.

%%%%%%%%%%%%%%
%%%%%%%%%%%%%%%%%%%%%%%%

\begin{figure}[ht]
\begin{center}
\includegraphics[width=10.2cm]{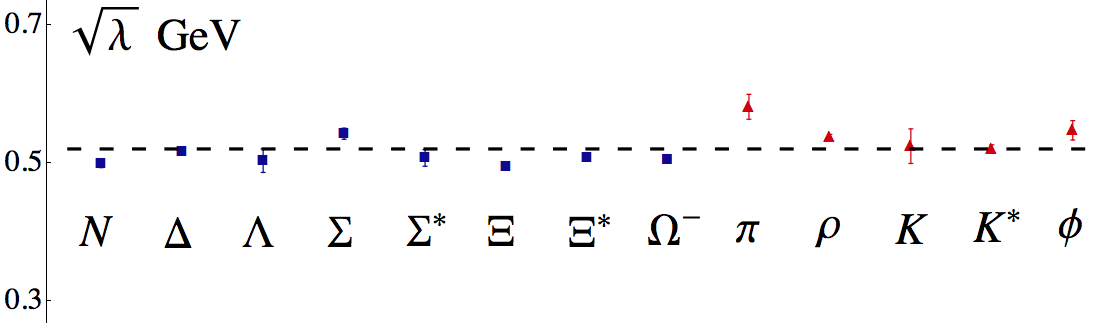}
\end{center}
\caption{\label{lightslope} Best fit for  the hadronic scale $\sqrt{\la}$ including all radial and orbital excitations of the light mesons and baryons.  The dotted line is the average value $\sqrt{\la}= 0.523$ GeV; it has the standard deviation  $\si =0.024$ GeV.}
\end{figure}

To include light quark masses in the computation of the hadron mass spectrum we leave the LF potential unchanged  as a first approximation and add the additional term of the invariant mass $ \Delta m^2 = \sum_{i=1}^n \frac{m_i^2}{x_i}$ to the LF kinetic energy. The resulting LF wave function is then  modified by  the factor $e^{-\frac{1}{2\la} {\Delta m^2}}$, thus providing a relativistically invariant form for the hadronic wave functions.  The effect of the nonzero quark masses for the squared hadron masses is then given by the expectation value of $ \Delta m^2$ evaluated using the modified wave functions~\cite{Brodsky:2014yha}. Using this procedure we can perform a consistent analysis of the excitation spectra of the $\pi, \rho, K, K^*$ and $\phi$   meson families as well as the $N, \Delta, \Lambda, \Sigma, \Sigma^*, \Xi$ and $\Xi^*$ families in the baryon sector to test the universality of the light hadron mass scale $\sqrt{\la}$. This is shown in Fig. \ref{lightslope}.

%%%%%%%%%%%%%%%%%%%%%%%
%%%%%%%%%%%%%%%%%%%%%%%

\subsection{Extension to the Heavy-Light Hadron Sector}

In the limit of zero quark masses the dynamical degrees of freedom, expressed through the LF variable $\ze$,  decouple from the the longitudinal momentum variables~\cite{deTeramond:2008ht}. In this domain the superconformal algebraic structure determines uniquely the superpotential $V(z) = \sqrt \la \, \ze$. Additionally, the non-trivial geometry of AdS space encodes all the kinematical aspects, whereas the modification of the AdS gravity action -described in terms of the dilaton profile $\varphi(z)$, includes the dynamics and determines the effective LF potential $U(\ze)$ for mesons in the LF bound-state equations~\cite{deTeramond:2013it}.  For heavy quark masses conformal symmetry is strongly broken and the form of the superpotential  and the dilaton are  unknown. However, if we can embed  the supersymmetric theory  in the modified AdS action, and the separation of the dynamical and kinematical variables also persist in the heavy-light domain, we can then derive uniquely the confinement potential for mesons and baryons in this domain. It turns out that this procedure reproduces the observed data to a reasonable accuracy and also allows us to make predictions for yet unobserved states~\cite{heavylight}.

\begin{figure*}[ht]
\centering
\sidecaption
\includegraphics[width=7.0cm]{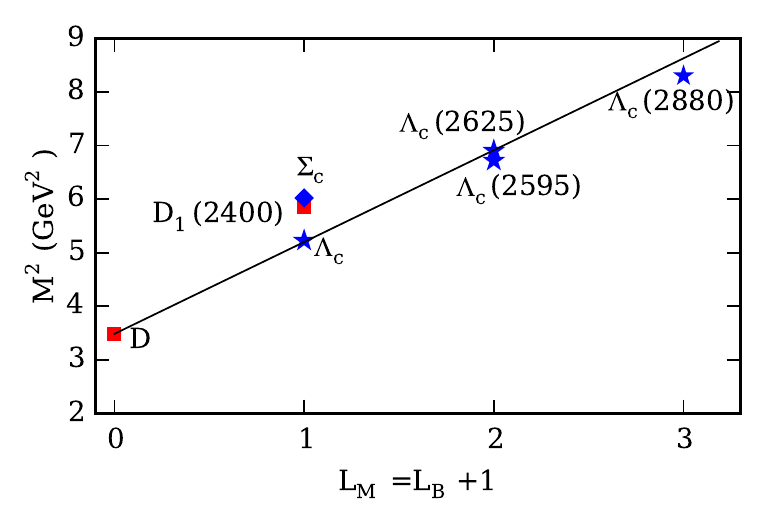}
\caption{Hybrid $D$ meson family and SUSY partner baryons with one charm quark. The data is from Ref. \cite{Olive:2016xmw}.}
\label{figcharm}
\end{figure*}

The actual embedding is carried out by equating the LF potential for mesons for an arbitrary superpotential $U_M$ \req{UM} with the expression for the potential given in terms of a dilaton profile \req{dilU} for $J = L_M$. The result is~\cite{heavylight}
 \beqa \label{dil}
 \vp(\ze) &=& \int d\ze\, \left(\la \ze\, \si(\ze) - \frac{ \la^2 \ze^2\,  \si'(\ze)}{\la^2 \ze^2 \, \si(\ze) + 2 (L_M-1) \la}\right), \\
 \label{pot}
V(\ze) &=& \frac{1}{2} \left(\la \ze \,\si(\ze) + \frac{ \la^2 \ze^2
\, \si'(\ze)}{\la^2 \ze^2 \, \si(\ze) + 2(L_M-1) \la}\right),
\enqa
where $\si(z)$ is an unknown function. However, since the dilaton profile must be independent of kinematical quantities, and hence  of the angular momentum $L_M$,  we must set the derivative $\si'(\zeta) =0$ in \req{dil} and \req{pot}. Thus we have $\si(\ze)= A$ with $A$ an arbitrary constant.  From \req{dil} and \req{pot}:
 \beq  \label{phiV}
  \vp(\ze) = \half  \la \, A \,\ze^2 + B,  \quad \quad 
  V(\ze) = \half \la \, A \, \ze. 
  \enq
Consequently, even for strongly broken conformal invariance the dilaton  has the same quadratic form obtained in the conformal limit. The constant $A$, however, is  arbitrary, and the strength of the potential is not determined.  We can set $B =0$ in \req{phiV} without modifying the equations of motion.

\begin{figure}
\centering
\sidecaption
\includegraphics*[width=9.0cm]{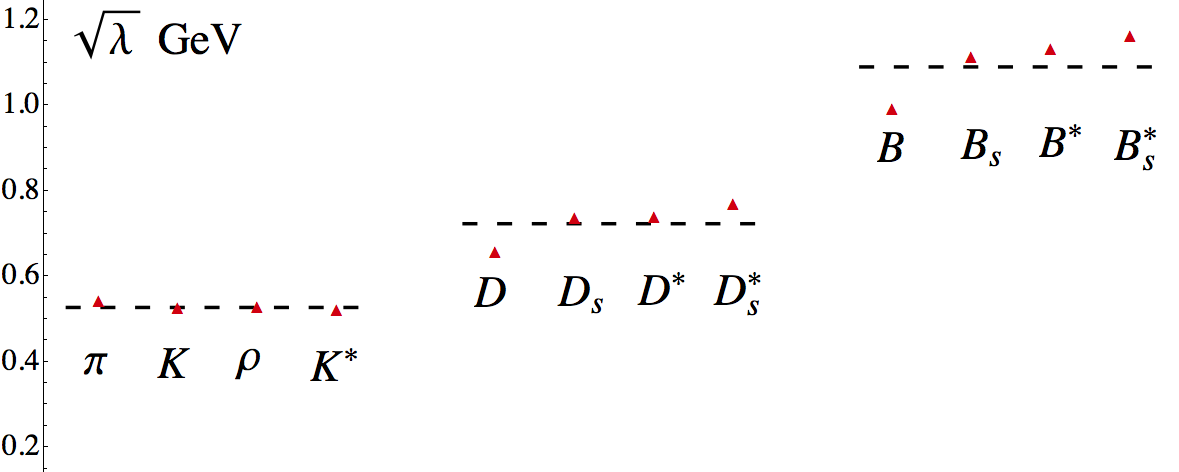}
\caption{Fitted values of $\sqrt{\la_Q}$ for different meson-baryon trajectories, as indicated by the lowest meson state on a given trajectory.}
\label{lambda-channel} 
\end{figure}

As an illustration we compare in Fig. \ref{figcharm}  the data  for the heavy-light  $D$ meson family and its SUSY partner baryon family containing one charm quark~\cite{heavylight}. The results presented in Fig. \ref{figcharm}  constitute a test of  the linearity of the trajectory predicted by the SUSY holographic embedding, and allows us to determine the dependence of the slope $\la_Q \equiv \half \la \, A$ on the heavy quark mass scale. The trajectory intercept is fixed by the lowest state.  In Fig. \ref{lambda-channel} the fitted values for $\sqrt{\la_Q}$ are presented for all the heavy-light trajectories~\cite{heavylight, Branz:2010ub}.  The results indicate a significantly greater dispersion of values of $\sqrt \la_Q$  for the case where the model is not constrained by conformal symmetry.

\section{Nucleon Electromagnetic Form Factors in Light-Front Holographic QCD}

We have recently performed a comprehensive analysis of the space-like nucleon electromagnetic form factors and their flavor decomposition in the framework of holographic QCD~\cite{Sufian:2016hwn}, where the specific modification of AdS space is given by the mapping to superconformal quantum mechanics in the light front. The covariant spin structure for the Dirac and Pauli electromagnetic nucleon form factors in the AdS$_5$ semiclassical gravity model incorporates the correct power-law scaling for a given twist from hard scattering and also leads to vector dominance at low energy.  For  twist $\tau$, the number of constituents $N$ in a bound state, the form factor (FF) is given by the analytic form~\cite{Brodsky:2007hb}
\beq \label{FFtau} 
  F_{\tau}(Q^2) =  \frac{1}{{\left(1 + \frac{Q^2}{M^2_{\rho_{n=0}}} \right)} 
 \left(1 + \frac{Q^2}{M^2_{\rho_{n=1}}} \right) \cdots  \left(1 + \frac{Q^2}{M^2_{\rho_{n = \tau -2}} } \right)},
 \enq
expressed as a product of $\tau -1$ poles along the vector meson Regge radial trajectory in terms of the $\rho$ vector meson mass $M_\rho$ and its radial excitations. Furthermore, the expression for the FF (\ref{FFtau}) contains a cluster decomposition: the hadronic FF factorizes into the $i = N - 1$ product of twist-two monopole FFs evaluated at different scales~\cite{deTeramond:2016pov} 
 \beq
F_{i }(Q^2) = F_{i = 2}\left(Q^2\right)  \,F_{i = 2} \left(\tfrac{1}{3}Q^2\right)\,  \cdots\,    F_{i = 2} \left(\tfrac{1}{ 2 i - 1} Q^2\right).
\enq
In the case of a nucleon, for example, the Dirac FF of the twist-3 valence quark-diquark state $F_1(Q^2)  = F_{i = 2}\left(Q^2\right)\,  F_{i=2}  \left(\tfrac{1}{3}Q^2\right)$ corresponds to the factorization of the proton FF as a product of a point-like quark and a diquark-cluster FF. The identical twist-3 expression from Eq. (\ref{FFtau}) is described by the product of two poles consistent with leading-twist scaling, $Q^4 F_1(Q^2) \sim  \text{const}$, at high momentum transfer.  The Pauli form factor $F_2$ is given instead by the $i = N + 1$ product of dipoles, and thus the leading-twist scaling $Q^6 F_2(Q^2) \sim  \text{const}$.

\begin{figure*}[htp]
\centering
\includegraphics[width=7.0cm]{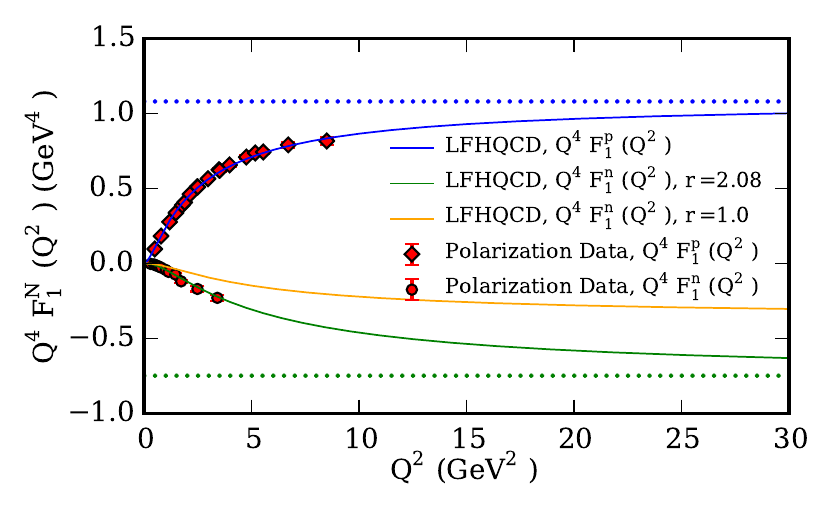}
\includegraphics[width=7.0cm]{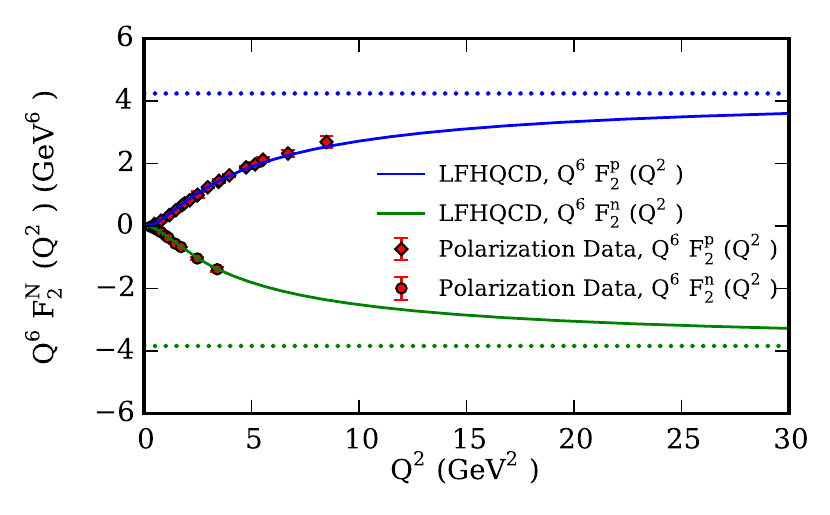}
\includegraphics[width=7.0cm]{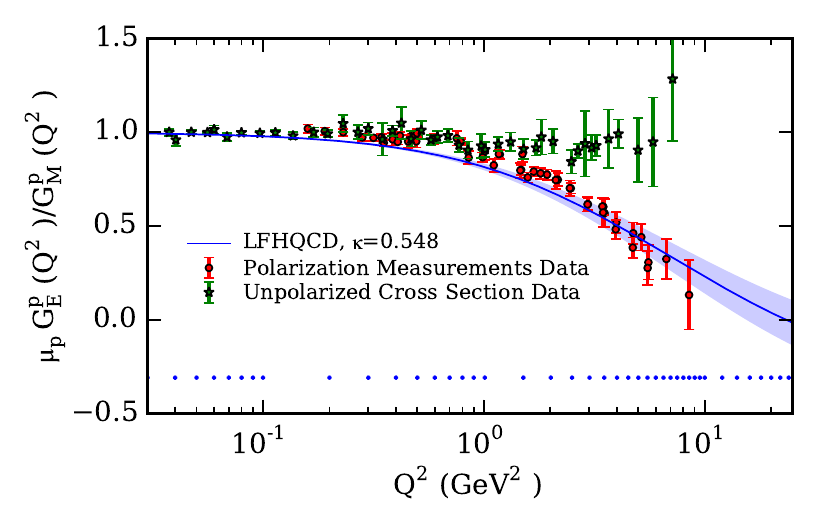}
\includegraphics[width=7.0cm]{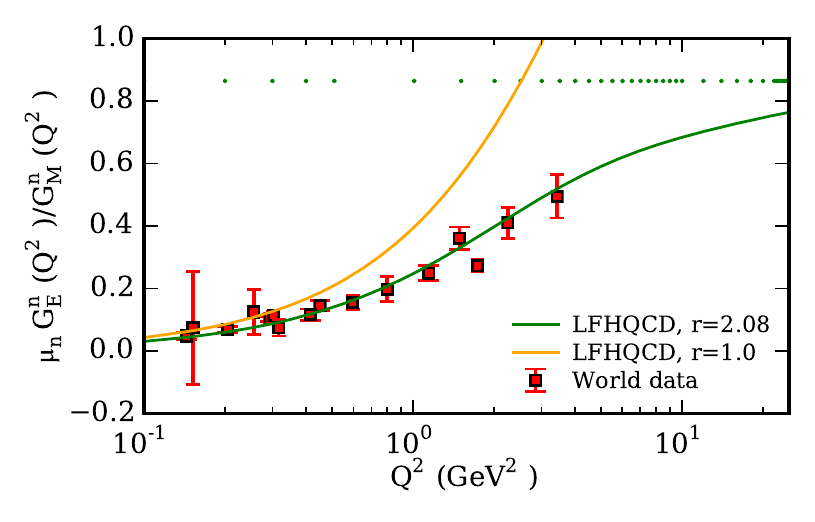}
\caption{Comparison of the holographic results with selected world data for the Dirac and Pauli form factors and  ratios of the electric to magnetic form factors for the proton and neutron. The orange  lines correspond to the SU(6) symmetry limit for the neutron form factor. The dotted lines are the asymptotic predictions. Reference to experimental data is given in~\cite{Sufian:2016hwn}. The band represents an estimated theoretical uncertainty of the model.}
\label{NFFs}
\end{figure*}

We have carried out our computations for any momentum transfer range, including  asymptotic predictions. Our results agree with the available experimental data with high accuracy as shown in Fig. \ref{NFFs}. Our results show that  the inclusion of the higher Fock components $\vert qqqq\bar{q} \rangle$ has a significant effect on the spin-flip elastic Pauli form factor (about 30$\%$ in the proton and about 40$\%$ in the neutron) and almost zero effect on the spin-conserving Dirac form factor. The free parameters needed to describe the experimental nucleon form factors are very few: two parameters for the probabilities of higher Fock states for the spin-flip form factors and a phenomenological parameter $r$,  to account for possible SU(6) spin-flavor symmetry breaking effects in the Dirac neutron FF, whereas the Pauli  form factors are normalized to the experimental values of the anomalous magnetic moments~\cite{Sufian:2016hwn}.

\section{Perturbative-Nonpertubative Interface in QCD and Determination of $\rm   \Lambda_s$}

The scale dependence of the coupling in the infrared (IR) domain is derived from the long-range confining forces. In holographic QCD it follows from the specific embedding of  light-front dynamics in AdS space~\cite{Brodsky:2010ur}. It is uniquely determined in terms of the dilaton profile which breaks conformal invariance, consistent with superconformal quantum mechanics~\cite{Fubini:1984hf}.  Likewise, the QCD coupling at short distances becomes scale dependent because of short-distance QCD quantum effects which are included in its definition: it is determined by perturbative QCD and its renormalization group equation.  The matching of the high and low energy regimes of the strong coupling  $\alpha_s$ determines the scale which sets the interface between perturbative and nonperturbative hadron dynamics~\cite{Deur:2016cxb}.

\begin{figure*}[ht]
\centering
\sidecaption
\includegraphics[width=5.8cm]{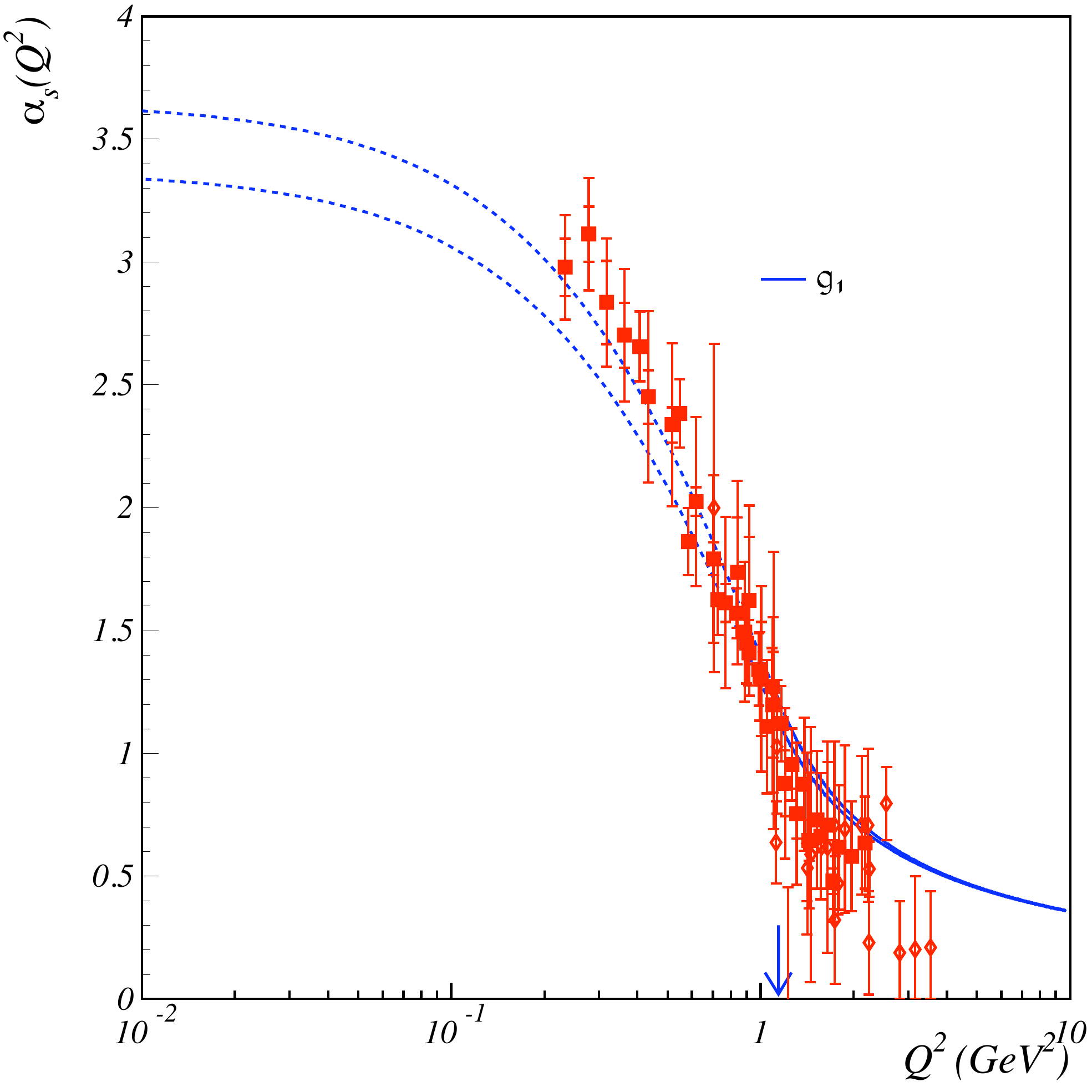}
\caption{Comparison between the experimental data and the strong coupling from holographic QCD. The continuous lines are the perturbative calculations. The dashed curves are the LF holographic QCD predictions matched to the non-perturbative domain. The transition  scale  $Q_0^2$ is shown  by the blue  arrow. The matching gives the value $\alpha^{IR}_s(0) = 3.51 \pm 0.62$ in good agreement with the value $\alpha^{IR}_s(0) = \pi$ from the specific definition of the effective charge $\alpha_{g1}$~\cite{Bjorken:1966jh}.}
\label{comp}
\end{figure*}

The strong coupling in the IR  has the analytic form~\cite{Brodsky:2010ur}
\beq
\alpha_s^{IR}(Q^2) = \alpha_s^{IR}(0) \, e^{- Q^2 / 4 \la},
\enq
which gives an accurate description of the effective charge $\alpha_{g1}(Q^2)$ determined  from the Bjorken sum rule~\cite{Bjorken:1966jh, Deur:2005cf}. We shall use the perturbative coupling $\alpha_{s}(Q^2)$ calculated up to order $\beta_{4}$ in the perturbative series of the $\beta$ function
\begin{equation} 
Q^{2}\frac{\partial\alpha_{s}}{\partial Q^{2}} =\beta\left(\alpha_{s}\right)=-\left(\frac{\alpha_{s}}{4\pi}\right)^{2}\sum_{n=0}\left(\frac{\alpha_{s}}{4\pi}\right)^{n}\beta_{n}, 
\end{equation}
which yields a five-loop expression for $\alpha_{s}$.  The matching of the high and low $Q^2$ regimes --equating $\alpha_S$ and its $\beta$ function, determines the scale $Q_0$ which sets the transition scale between the perturbative and nonperturbative regimes.  This is illustrated in the $g_1$ scheme in Fig. \ref{comp}.  One can also compute the perturbative QCD scale $\Lambda_s$ in any scheme from the universal hadronic scale $\sqrt \la$. Using the value $\sqrt \la = 0.523 \pm 0.024$ GeV determined from the light hadron spectrum~\cite{Brodsky:2016yod} (See Fig. \ref{lightslope}) one finds $\La_{\overline{MS}} = 0.339 \pm 0.019$  GeV for $n_f = 3$, in excellent agreement with the world average, $\La_{\overline{MS}} = 0.332 \pm 0.019$ GeV~\cite{Olive:2016xmw}.

\section{Concluding Remarks}

Superconformal quantum mechanics and its light-front holographic embedding in a higher dimensional gravity theory provides a viable new framework which incorporates key nonperturbative aspects of hadronic physics that are not obvious from the QCD Lagrangian, including the emergence of a universal mass scale  and color confinement out of a nominally conformal theory.  The pion is massless in the limit of zero quark masses, and it has no supersymmetric baryon partner. The bound-state equations depend explicitly on the orbital angular momentum,  and thus chiral symmetry is broken from the outset in the Regge excitation spectrum.  Furthermore, this new approach gives unexpected connections between mesons and baryons  and allows  the description of physical observables, such as  meson and nucleon structure functions, transverse momentum distributions,  and distribution amplitudes, which are not amenable to a first-principle computation. In this article we have briefly described new developments, such as the extension of the framework to actually compute the heavy-light spectrum.  This implies linear trajectories not only for light hadrons, but also for hybrid ones. Indeed, existing data are not in contradiction to the linearity predicted by the supersymmetric embedding~\cite{heavylight}. Other recent applications include the description of nucleon form factors, with predictions over an extensive energy range  in view of the upcoming experiments at JLab~\cite{Sufian:2016hwn}. We have also discussed  the behavior of the QCD coupling in the infrared region, and its matching with the short-distance regime, which allows the prediction of the perturbative QCD scale $\Lambda_s$  and the transition scale $Q_0$ from the perturbative to the nonpertubative domains~\cite{Deur:2016cxb}. The underlying superconformal symmetry provides strong constraints on the embedding geometry and the resulting confinement dynamics. It also provides convenient nonperturbative boundary expressions at the hadronic scale which are then driven by QCD evolution equations to higher energy scales.

\section*{Acknowledgments}

GdT wants to thank the organizers of Confinement 12 for their generous hospitality at Thessaloniki. S.J.B. is supported by the Department of Energy, contract DE--AC02--76SF00515. SLAC-PUB--16869. A. D. is supported by the U.S. Department of Energy  Office of Science and Office of Nuclear Physics under contract DE--AC05--06OR23177.

\end{document}